%% file: IRAS1654.tex
\documentclass[12pt,preprint]{aastex}
\usepackage{t1enc}
\usepackage{graphicx}
\usepackage{epstopdf}
\usepackage{natbib}
\usepackage{lscape}
\newcommand{\kms} {km~s$^{-1}$}
\newcommand{\iras}{IRAS 16547-4247}

\newcommand{\snu}{$S_\mathrm{\nu}$}
\newcommand{\lsun}{$L_{\odot}$}
\newcommand{\msun}{$M_{\odot}$}

\newcommand{\jpb}   {$\rm Jy~beam^{-1}$}

\newcommand{\gap}%
{\raisebox{-0.5ex}{$\stackrel{\scriptstyle >}{\scriptstyle \sim}$}}

\begin{document}

\shorttitle{Compact sources in UCHII regions}
\shortauthors{Masqu\'e et al.}

\title{GMRT observations of the protostellar jet associated with \iras }

\author{
Josep M. Masqu\'e\altaffilmark{1},
Solai Jeyakumar\altaffilmark{1},
Miguel A. Trinidad\altaffilmark{1},
Tatiana Rodr\'iguez-Esnard\altaffilmark{1}, \&
CH Ishwara-Chandra\altaffilmark{2}
} 

\altaffiltext{1}{Departamento de Astronom\'ia, Universidad de Guanajuato, Apdo. Postal 144, 36000 Guanajuato, M\'exico}
\altaffiltext{2}{National Centre for Radio Astrophysics, TIFR, Post Bag No. 3, Ganeshkhind Post, 411007 Pune, India}
\begin{abstract}

We report continuum GMRT observations aimed to explore the behavior of the
jet associated with IRAS 16547-4247 at very low frequencies (325 MHz and
610 MHz). The obtained maps reveal an elongated morphology in the SE-NW direction.  In addition, the 610 MHz map shows three knots associated to the elongated morphology that seems to correspond to the triple radio source identified as the jet seen at higher frequencies. However, at 325 MHz, although an elongated morphology is also observed, only two knots appear in the map. By
comparing our knot positions  at both frequencies with a precessing jet model used in a
previous work, we find that the knots fall closely to the wiggling path
and, hence, they likely represent shocks of the material of the precessing 
jet with the medium. Only the nature of the southernmost knot detected
at 325 MHz remains unclear. Besides, we found that the whole emission of the lobes
is non-thermal down to very low frequencies and that the possible associated emission mechanisms work differently in in both lobes, causing the discrepancies between  observed frequencies. To explain these discrepancies we investigate mechanisms such as synchrotron radiative losses in a magnetic field of $\sim$0.5 mG in the shocks and find this possibility unlikely to occur. Alternatively, we invoke interaction with an inhomogeneous medium as the most probable scenario.

\end{abstract}
  

\section{Introduction}

Collimated winds constitute an ubiquitous phenomena in the star forming process. It is believed that they are launched at high velocities via magnetohydrodynamic proceses occurring in an accretion disk surrounding a young protostar  \citep{shu1987,ouyed1997,fendt2006,zanni2007}. In a wide range of scales, collimated winds give rise to several observational signposts  \citep[see][]{bally2016}. At scales ranging from tens to hundreds of AU from the protostar, the wind is partially ionized and its radio emission is dominated by free-free radiation\citep{reynolds1986, anglada1996a, anglada2018}. When angularly resolved, the winds appear as elongated structures associated with the protostar and are called 'thermal radio jets' \citep[e.g.][]{rodriguez1990,girart1996}. On larger scales, the interaction between the collimated wind and the surrounding cloud is detected. Such interaction can be observed as molecular outflows that are produced by the molecular material of the cloud entrained by the collimated wind \citep{raga1993,raga2003, anglada1995}, or with strong shocks between the high velocity outflowing gas and the interestelar medium. At optical wavelengths, Herbig-Haro (HH) objects are manifestations of these powerful shocks \citep[][]{reipurth2001}. Additionally, the high excitation of HH objets makes them emitting objects at other frequency bands \citep[e.g. HH 80-81:][]{rodriguez1989a,lopezsantiago2013}, including atomic and molecular lines \citep[e.g.][]{eisloffel2000, anglada2007, lopez2009}.  Large scale maps of star forming regions show that HH objects can be usually associated in chains of knots forming large HH systems ending up to a bow shock, which is associated with the terminal working surface of the jet \citep[see][]{reipurth2001}.

Despite the wide range of stellar masses determined for young objects, it appears that all of them have associated jets suggesting, not without controversy, that high mass protostars form under a similar mechanism than low mass protostars. However, massive protostars undergo a much intense accretion phase than low mass stars \citep{walmsley1995,pudritz2007}. If the protostellar accretion is intimately bound to the launch of collimated winds, then much powerful jets should be observed in high mass star forming regions with respect to their low mass counterparts. A clear example is found in the HH 80-81 system for which projected sizes of 10.3 pc \citep{masque2012b} and tangential velocities between 200 and 1000 \kms\ have been measured \citep{marti1995, marti1998, masque2015}. The morphological resemblance of this jet emanating from a high mass protostar to other jets found in low mass young objects suggests that protostellar jets scale their properties depending on the mass of the powering source.  This scenario is strengthened by recent observations that resolve the disk around the massive star powering the HH 80-81 radio jet \citep{girart2018}.      

However, the shocks observed in massive star forming regions can have different properties from those observed around low mass young objects, owing to the higher velocities in which the jet material slams on the ambient medium. Up to the date, a handful of massive YSO have been reported to present
negative spectral indices at radio wavelengths (e.g. HH 80-81: \citealt{marti1993, masque2012b}; L778: \citealt{girart2002a}, W3 (OH):   \citealt{wilner1999}). This implies a significant contribution of non-thermal emission in the shocks that is usually proposed to be synchrotron
emission produced by a population of ultra-relativistic electrons. In this sense, \citet{carrascogonzalez2010} performed successfully a direct measure of the magnetic field strength and direction in the HH 80-81 jet, providing for the first time, evidence of the presence of a population of relativistic particles trapped in a magnetic field present in the jet. 
Recently, \citet{ainsworth2014} reported for the first time a negative spectral index in a bow shock associated to an outflow emanating from a low-mass YSO, suggesting that relativistic particles are also present in low-power jets.

Relativistic populations of particles are easily found in the AGN jets \citep[e.g.][]{blandford1982} 
and supernova remnants  \citep[e.g.][]{castro2010}, where gas flows reach velocities up to a significant fraction of the speed of light. However, the material of protostellar jets moves at a speed of few hundred of \kms. 
For theses velocities, particle acceleration mechanisms have been proposed to work on the free electrons of the gas, such as strong shocks created by the jet in the ambient 
medium.  In order to understand the acceleration process, detailed simulations of 
propagating relativistic electron beams were carried out 
using numerical solutions to the Fokker-Planck equation \citep{casillasperez2016}. 
These simulations suggest that the electron population will quickly lose most 
of the low energy electrons due to collisions with the ambient gas. This provides some diagnostics to be tested, especially using low frequency observations around massive young stellar objects where strong shocks with its surrounding medium are expected to be found as a consequence of protostellar activity.

\iras, with a luminosity of $6.2 \times 10^4$ \lsun, is
associated with a massive star forming region located at a distance of 2.9
kpc. Continuum and molecular line observations at mm wavelengths show \iras\ coincident with a dusty dense core of 10$^3$ \msun\ \citep{garay2003}. The same authors carried out radio observations over the region
and found three components aligned in the SW direction with the outer
components symmetrically separated from the central source by 10$''$ (0.14
pc in the plane of sky). The triple radiosource was proposed to constitute
a central jet with two outer lobes representing the working surfaces of
the jet with the medium. Interestingly, the spectral indices of the lobes ($\alpha=-0.61$ and $-0.33$, where $\alpha$ is defined as \snu $\propto \nu^{\alpha}$ and is 
derived from 1 to 10 GHz, approximately) 
suggest the presence of non-thermal emission. Besides, \citet{brooks2003},
through IR observations, show that the collimation of the outflow persists up
to 1.5 pc from the powering source, making \iras\ the first case
of a collimated outflow powered by an O-type star. 
A further study over the region using molecular lines reveals an energetic outflow associated to \iras, making evident the powerful ejecta form the massive star \citep{garay2007}. However, \citet{rodriguez2008}, using two sets
of 3.6~cm
observations, do not detected proper motions along the axis of the outflow in the
outer lobes, 
although they show evidence of precession.

IRAS 16547-4247 has also been proposed as a potential high-energy source by
\citet{munaradrover2013}, who using XMM-Newton observations, found that  IRAS 16547-4247 is a
hard X-ray source. They also found that may be difficult to explain the X-ray emission as
non-thermal emission,
but rather as thermal Bremsstrahlung plus photo-electric absorption in the cloud. 
At smaller scales close to the powering source, 
0.85~mm continuum observations revealed that  IRAS 16547-4247 is split into two
compact sources, 
one of them spatially associated with  IRAS 16547-4247, and the other one to the
west \citep{zapata2015}. 
In this way, \citet{higuchi2015}, using $^{12}$CO(3-2) observations, detected two
outflows, one of them aligned
in the NW-SE direction, and the other one in the NE-SW direction, which suggest
multiple driving sources
powering outflows in the region.

In this paper we report GMRT observations of the jet associated with \iras, aimed to explore for the first time the behavior of its associated shocks at very low frequencies, where effects such as synchrotron self-absorption and the low energy cut-off of the electron spectrum of the present electron population become important. Very low frequency studies (i.e. $<1$ GHz) over proto-stellar jets are not frequent because, in addition to the low angular resolution (i.e. resulting in a poor astrometry) and terrestrial radio interference, contamination from the galactic background is notable. 
The observations and data reduction, including actions performed to overcome the mentioned difficulties, are described in Sect. 2 and in Sect. 3 we report the results. A discussion about our tentative findings is shown in Sect. 4.  Finally, in Sect. 5 we summarize our work.


\section{Observations}

The GMRT observations were conducted in two observing blocks on 2017 Feb 21th and 28th. 3C286 was used as primary and bandpass calibrator, while  J1714-252 was used for the gain calibration.
The receivers were tuned at the 325 MHz band and 610 MHz band during the first and second day, respectively. The details of the observations are given in Table \ref{observations}.

The GMRT data were flagged, calibrated and imaged using the SPAM pipeline \citep{intema2017} which
includes several rounds of self-calibration and iterative flagging. Since the target is near the galactic plane, correction
to the system temperature $T_\mathrm{sys}$ arising due to excess background was needed.
This correction was derived using the all-sky map at 408 MHz by
\citet{haslam1995} as part of SPAM pipeline.
At 325 MHz, the Tsys correction factor applied was 6.10 and at 610 MHz, the
$T_\mathrm{sys}$ correction factor was 2.21.

At low frequencies, the ionosphere introduces some propagation delay differences between the elements in the array.  A method widely used to correct for these phase errors
consists in performing \emph{self-calibration} using several bright and compact sources in the initial image \citep{pearson1984}.  An improved version of this scheme has been employed in the SPAM pipeline where the ionospheric corrections have been estimated for several directions. This scheme has been shown to produce significant improvement in
image fidelity as compared to standard self-calibration \citep{intema2009}. Though the self-calibration is well constrained, it could introduce small astrometric errors. 

\begin{figure}[thbp]

\resizebox{1.1\textwidth}{!}{\includegraphics[angle=0]{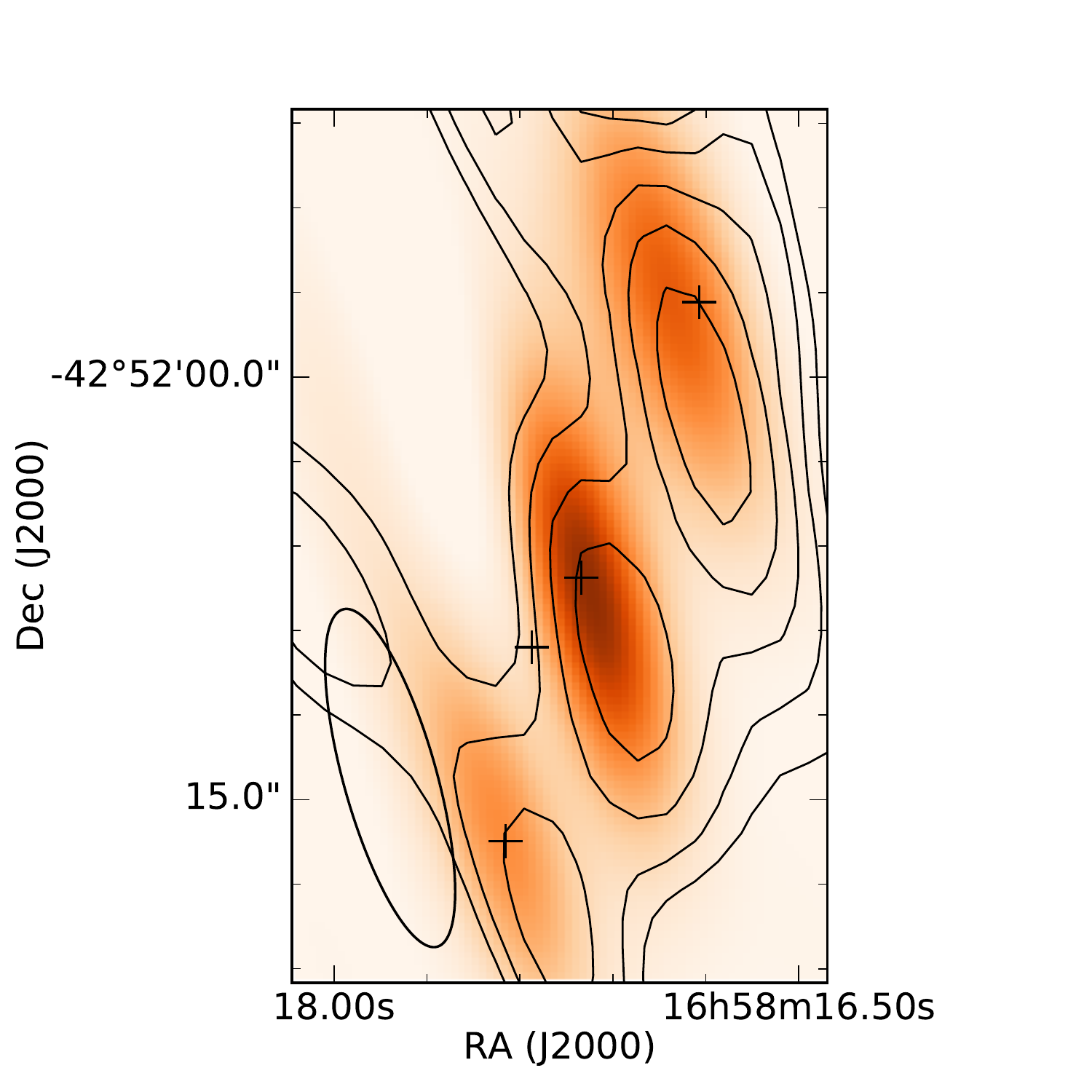}}
\caption{Comparison of the map obtained at 610 MHz (contours) before performing the astrometric correction (see text) with 
the 4.8 GHz data of \citet{rodriguez2005} convolved to the GMRT 610 MHz beam size (color scale).  Contours are $C$ times $-2^{0}, 2^{0}, 2^{1/2}, 2^{1}, 2^{3/2}$ and $2^{2} $, where $C$ is given by 2.5 times the $rms$ noise of the map (0.3 m\jpb\ ). The beam is shown in the bottom left corner. Although we corrected the astrometry of our data using field sources, this figure is shown to illustrate the displacement of our maps with respect other observations and how our offset corrections are adequate. \label{comparisson}}
\end{figure}

The astrometry of our data was checked using field sources from the GPSW catalog \citep{white2005}. From this catalogue, we employed sources 343.021+0.312, 343.144+0.032, 343.260-0.002 and 343.210-0.191 to obtain average offset corrections that were applied to our maps. In the source selection we intended to avoid multiplicity or extended emission.  Our results are 0.4 and 3.7 arcsec to the west and north, respectively, for the 325 MHz map, while for the 610 map the offsets are 1.2 and  5.1 arcsec to the east and north, respectively. The standard deviation of the corrections  was  $\sim 1$ arcsec. This uncertainty was added to the 10\% of the beam size as the typical positional accuracy of GMRT. Furthermore, we checked these offsets by comparing our 610 MHz map (before correction) with 
high frequency images at 4.8 GHz \citep{rodriguez2005} convolved to our 610 MHz beam and found consistent results. 
This can be assessed in Figure \ref{comparisson}, where the knots at 610 MHz appear displaced with respect to the  4.8 GHz knots with an offset similar to that derived above. 
Because the emission peaks may slightly differ at low and high frequencies due to the contribution of different emission mechanisms, we adopted the offsets derived with point sources as the most accurate ones. After shifting the maps with the appropriate offsets, the sources belonging in the jet seen in our maps fall approximately in the jet axis derived in previous works.

\section{Results}

\subsection{Emission maps}

\begin{figure}[thbp]

\resizebox{1.1\textwidth}{!}{\includegraphics[angle=0]{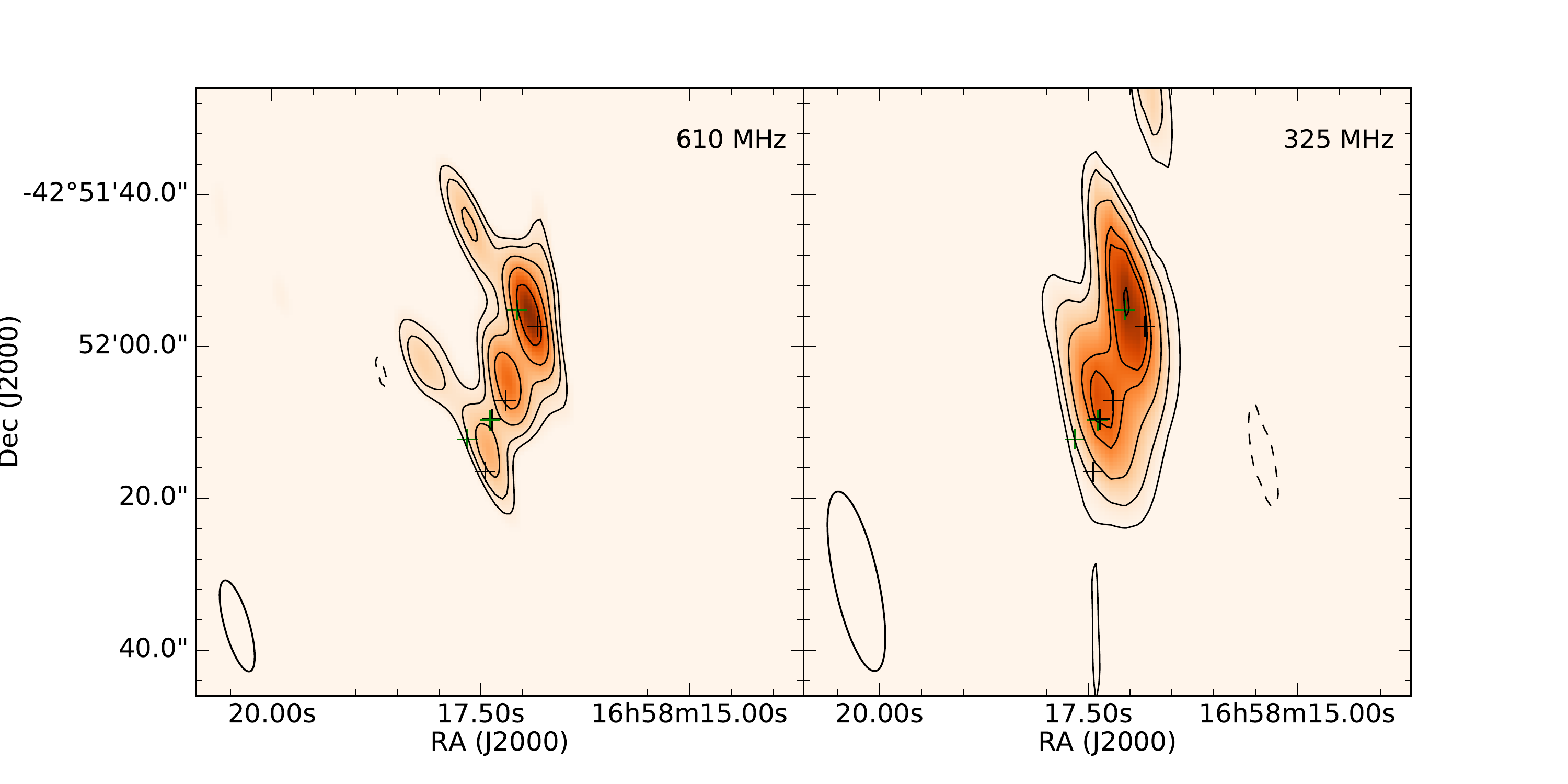}}
\caption{
GMRT continuum emission maps at 325 MHz (right panel) and 610 MHz (left panel) of \iras\ (color and contours) after the astrometric correction. Contours are $C$ times $-2^{0}, 2^{0}, 2^{1/2}, 2^{1}, 2^{3/2},2^{2} $ and $ 2^{5/2}$, where $C$ is given by 2.5 times the $rms$ noise of each map (0.6 m\jpb\ for 325 MHz and 0.3 m\jpb\ for 610 MHz). 
Black crosses represent the  peak position of the  knots derived in \citet{garay2003}, while 
cyan crosses represent the sources sources that are outside of the jet path derived in \citet{rodriguez2005}. The beam is shown in the bottom left corner. \label{cont_maps}}
 \end{figure}

  Fig. \ref{cont_maps} shows the continuum emission of \iras\ at low radio frequencies. From these maps, the jet structure becomes evident as an elongated morphology with a size of $\sim30'' \times 20''$ in the SE-NW direction.

   At 610 MHz, the jet shows three aligned continuum knots representing the triple radio source first detected by \citet{garay2003}.  The northern knot is the brightest and the southern knot is the weakest and both sources are symmetrically separated $\sim10''$ from the central source in opposite directions, in agreement with \citet{garay2003}. Two additional sources of $\sim$1 mJy appear in the map
about $15''$ East to the jet. While the southern source is located too far away to be considered related to the jet, the northern source could belong to the jet of we account for a wiggling motion. However, both sources are weak and were not reported before. Thus, they will not be further discussed in this work. 

 At 325 MHz, an elongated structure with two continuum knots separated $\sim15''$ is revealed. While the location of the northern knot is nearly coincident with the position of the northern peak at 610 MHz, the southern knot can not be convincingly associated with any of the 610 MHz counterparts. Similarly as for the 610 MHz data, the northern knot seen at 325 MHz is the brightest.

Figure \ref{cont_maps} also shows the position of the peaks reported in \citet{garay2003} using observations with similar angular resolution than our observations at 625 MHz band. We measured the distance between the peak position of the northern and southern knots separately for the 325 MHz and 610 MHz bands. For a more exhaustive comparison, we performed the same measurement between the northern and southern peaks reported in \citet{garay2003}. We take the distance between the northern and southern peaks as representative of the jet size at each frequency. We obtained the following results:  $20 \pm 0.1''$ \citep[8.6 GHz, ][]{garay2003}, $19 \pm 1''$ (610 MHz) and $15 \pm 2''$ (325 MHz), whose uncertainties are estimated from the signal to noise ratio and beam size. Although the discrepancy in separation between knots at 8.6 GHz and 610 MHz is not significant, this separation estimated at 325 MHz appears much shorter than in other bands. The derived sequence could suggest that the separation between the emission of the terminal shocks of the jet decreases at low frequencies.

 The angular resolution of our 610 MHz map is similar to that of the 1.4 GHz map of \citet{garay2003}. However, the appearance of the latter map is different: the jet shows a northern peak with a tail extended to the south. The most plausible explanation on this relies on the position angle  of the beam of the 610 MHz map, that makes possible to resolve better the jet along its direction than the 1.4 GHz map.  To corroborate this, we degraded the angular resolution toward the direction of the jet in the 610 MHz map and obtained the same jet appearance seen in the 1.4 GHz map. On the other hand, in spite of the lower angular resolution of the 325 MHz map with respect to the 610 MHz map, the former shows a southern knot associated with the jet not seen at higher frequencies. According to the trend of the jet structure with frequency found above, this knot is not expected to be prominently detected if the emission in the jet behaves similarly at all frequencies. Thus, the 325 MHz emission appears to arise from a different spot than the southern tip of the jet seen at higher frequencies.

\subsection{Determination of spectral indices}

To extract fluxes from the maps we fitted to the jet emission three and two Gaussian models simultaneously for the 610 MHz and 325 MHz maps, respectively,  with the CASA function $fitcomponents$ from the CASA image toolkit. In order to prevent contamination of thermal emission from the central jet to the flux of the lobes at 325 MHz, a synthetic Gaussian model with the beam size at this frequency was subtracted to the corresponding position before performing the fits in the map. The thermal flux of the Gaussian model (1.3 mJy) was estimated from the fit of the central panel of Figure \ref{spectra} that coincides with the $3-\sigma$ upper level for the 325 MHz flux.  The flux values obtained from the Gaussian fits are given in Table \ref{sources_tot}, together with the peak positions and the deconvolved source sizes. The central knot is unresolved indicating us that the very low frequency emission is not extended towards the central source, whereas the northern and southern knots have slightly extended sizes ($\sim11'' \times 4''$). Nevertheless, these results are marginal and probably due to the fact that the lobes of the jet are composed of several smaller components as revealed by high angular resolution observations \citep{rodriguez2005,rodriguez2008}.

Figure \ref{spectra} shows the fit to the observed fluxes reported in the present work and in previous papers. In this fit we assumed that the emission have a power law dependence with frequency. From the literature, we selected data of \citet{garay2003} as they employed observations with similar angular resolution than ours. The numerical results of these fits are reported in Table \ref{espectra}.  As seen in the table, the total fit is obtained under a reasonably good correlation coefficient and our slope values do not differ significantly from those reported before  \citep{garay2003,rodriguez2005}. 

The dispersion seen in some points in the graphs is not crucial to corrupt our fit results and occurs often when comparing data with a wide time range. Flux density discrepancies between different observations can be attributed to differences in instrumental calibration or in $uv$-coverage. Otherwise, they can be due to flux variability inherent to thermal jets and HH knots, as they can show significant bright changes on timescales of few years \citep{rodriguez2000, rodriguez2001,galvanmadrid2004}. For instance, flux measurements for some of the knots of the HH 80/81/80N jet have are discrepancies up to 50\% between observations of different epochs \citep{marti1993, masque2012b, masque2015}. In \iras, some of the radio sources in the region show flux variations up to 40\%  \citep{rodriguez2008}. For the particular case of the southern lobe, an increase with more than a factor of two in the 4.8 GHz flux with respect to \citet{garay2003} is observed, even though in this case we are comparing observations with different $uv$-coverage and, probably, the uncertainties are notable.

\section{Discussion}

\subsection{Precession of the jet}

The angular separation of the northern and southern knots with the central source is similar ($\sim10''$), suggesting that the shocks that produce them can be associated to the same ejection event. However, as seen above, 
there are some differences between the northern and southern knots. First, the northern lobe is significantly brighter than the southern lobe. According to the inclination of the outflow axis from the line of sight of 84 degrees, with the red-shifted lobe pointing to the north, derived in \citet{garay2007} the greater brightness at low radio frequencies observed for the corresponding knot is probably intrinsic. Second, as shown in Fig \ref{cont_maps}, our derived peak position at 325 MHz for the southern knot is not fully coincident with the southern knot position seen at higher frequencies \citep[e.g.,][]{rodriguez2005}. 

It is interesting to note that the continuum peak position of the components of the jet are consistent between frequencies greater than 325 MHz as observed  by comparing between several works \citep{garay2003,rodriguez2005,rodriguez2008}  and our 610 MHz map. This indicates us that their associated emission arises from the same shocks and that the knots lack for proper motions. On the other hand, the position of our knots at 325 MHz shows evident discrepancies with these studies, suggesting that the very low frequency data make evident different characteristics in the shocks associated with the jet lobes. The peak position of the three knots detected at 610 MHz (see Table \ref{sources_tot}) are aligned with a P.A. of $\sim70\arcdeg$, similar to that of \citet{rodriguez2005}. On the other hand, at 325 MHz, the position of the northern knot has a position angle of $\sim80\arcdeg$, still consistent with being part of the jet within uncertainties ($\sim10\arcdeg$), but the southern knot has a P.A. of $\sim30\arcdeg$, which is similar to the uncertainty in P.A. for this knot. This rises the question whether this knot belongs to the jet or not.

\citet{higuchi2015} found evidence for at least two intersecting outflows emanating from \iras. Supported by the distribution of Class I methanol masers \citep{voronkov2006}, they estimated the directions of the outflows as shown in their Figure 5. The location of the southern knot at 325 MHz agrees with being part of the northeast-southwest outflow.  
However,  the lack of additional radio counterparts for this outflow, added to the large uncertainty for our derived PA for the southern knot at 325 MHz, makes this possibility unlikely.

Alternatively, the detected knots at very low frequencies could represent other components of the jet reported in \citet{rodriguez2005}, which are not perfectly aligned due to the jet precession \citep{rodriguez2008}. In Figure \ref{jet_maps}, we show our maps with all the components of the jet marked with crosses and a precession model for the jet superimposed. The uncertainties in the alignment process added to the error in the position peak determination in the Gaussian fits are indicated in the figure.  From this figure, we can see that the position of the knots reported in this paper falls approximately within the jet path and, hence, they are possibly tracing shocks as the precessing jet digs into the cloud. In this scenario, the 325 MHz emission would probe shocks with different properties along the jet. Furthermore, the somewhat curved morphology of the knots, suggests that even at very low frequencies, hints of precession can be inferred.

  \subsection{Non-thermal emission from the lobes of the jet} 
    
\begin{figure}[thbp]
\resizebox{0.8\textwidth}{!}{\includegraphics[angle=0]{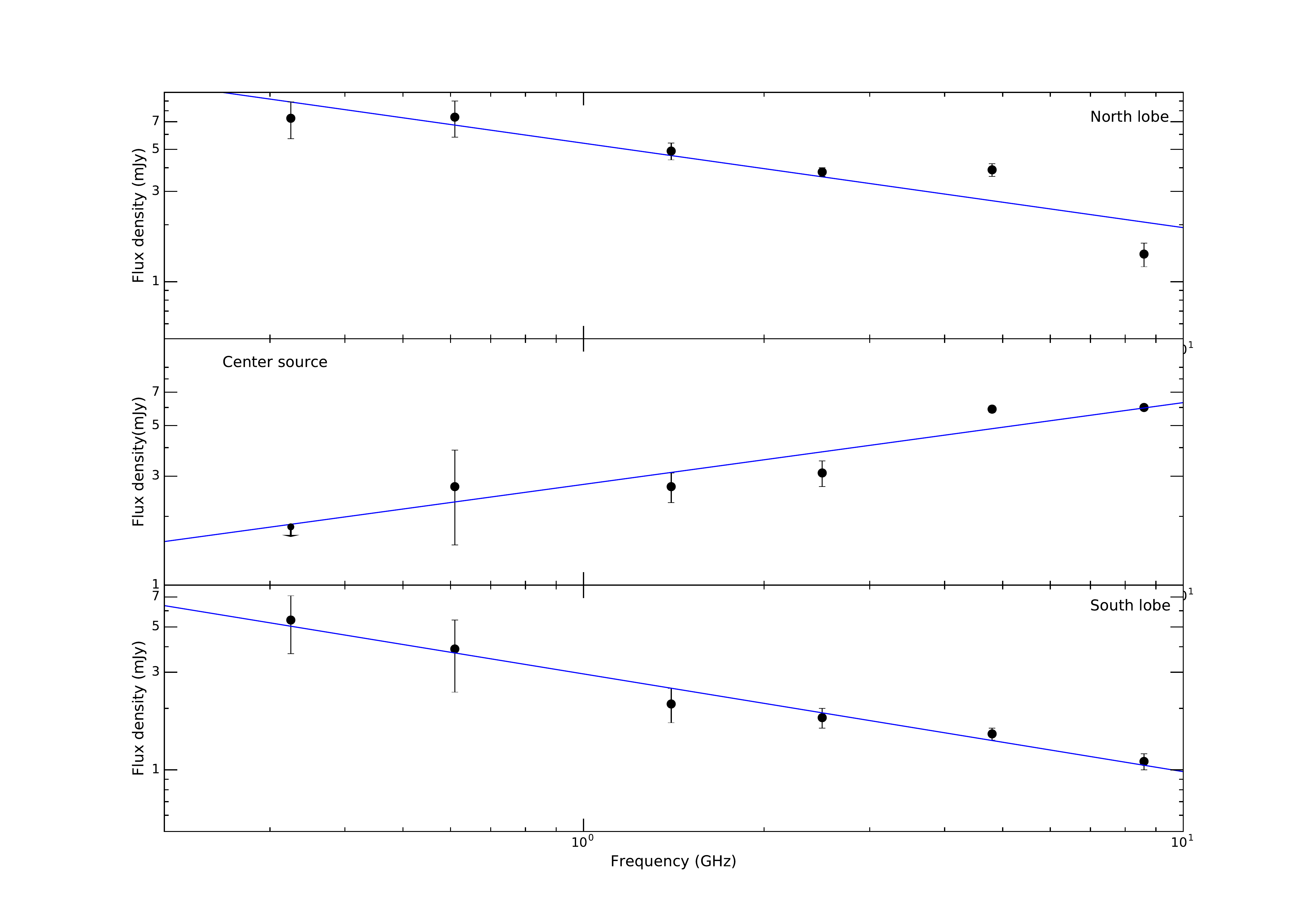}}
 \caption{Plots of the flux density vs frequency of  the northern lobe (top), central source (middle) and southern lobe (bottom) using the observed data  \citet{garay2003} and ours (black dots and error bars). The blue solid line shows the best fit to the observed fluxes assuming a power law dependence with the frequency. The arrow in the middle panel at 325 MHz represents 3 times the $rms$ noise level.   \label{spectra}}
\end{figure}

\begin{figure}[thbp]
\resizebox{1.1\textwidth}{!}{\includegraphics[angle=0]{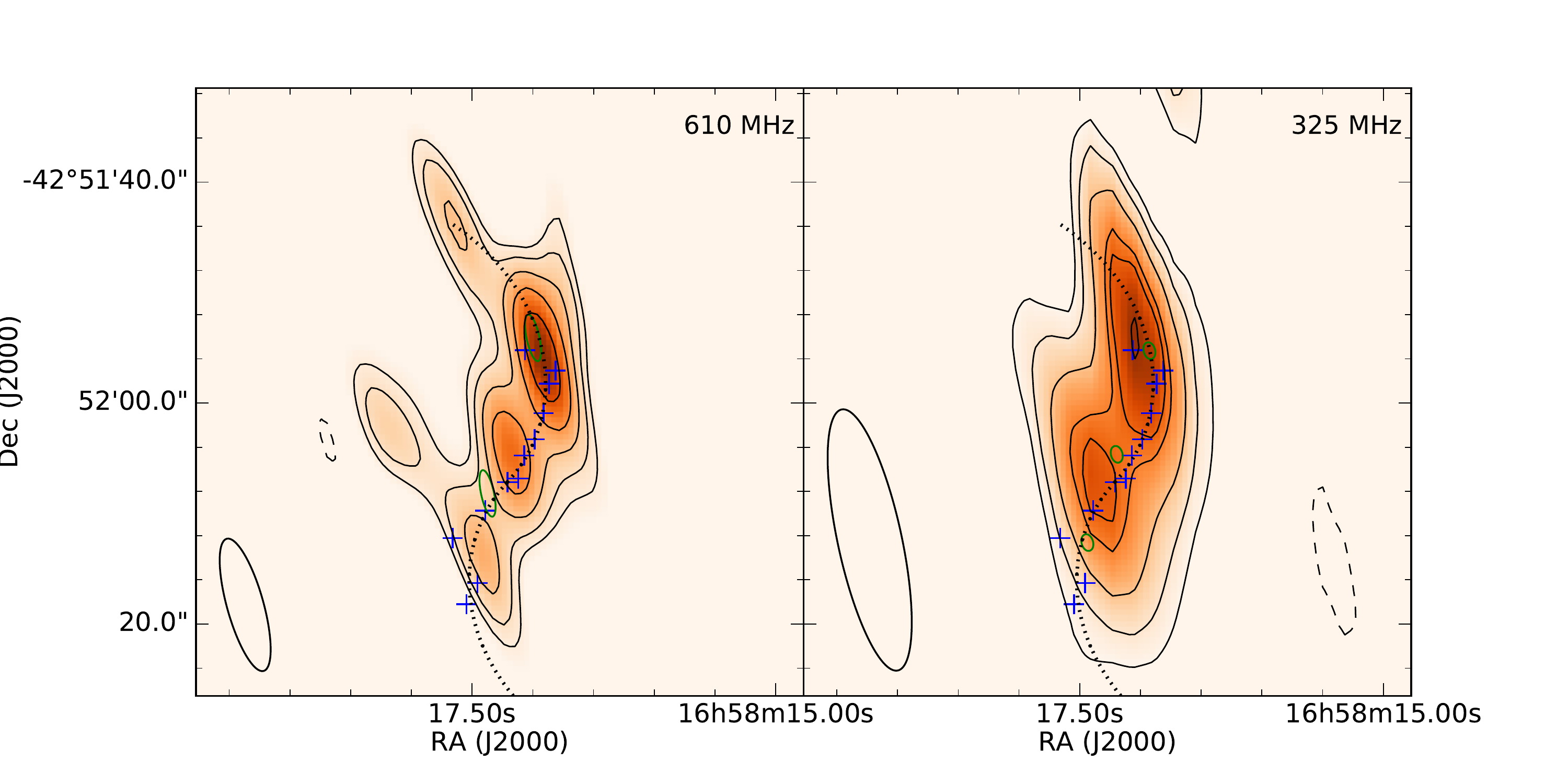}}
\caption{
GMRT continuum emission maps at 325 MHz (right panel) and 610 MHz (left panel) of \iras\ (color and contours). Contours are the same as in Fig. \ref{cont_maps}. The blue crosses represent the peak positions of all the outflow sources observed in 2006 \citep{rodriguez2008}. The green circles represent the peaks resulting from Gaussian fits to the data and show the uncertainties in the alignment process (see text). In order to compare the position of the knots at both frequencies, we show the ellipses corresponding to the 325 MHz peaks in the 610 MHz map and the other way around for the 325 MHz map. The dashed line represents the spiral model employed in \citet{rodriguez2008}. \label{jet_maps}}
 \end{figure}

     In Figure \ref{spectra} we show the fits to the flux density of the knots of the jet reported in Table \ref{sources_tot} assuming that they follow a power law dependence with frequency. We also considered that the emission of the southern 325 MHz knot belongs to the southern lobe. Our results are in agreement with previous values obtained with a narrow bandwidth \citep{garay2003,rodriguez2005} as indicated by the good correlation coefficient seen in Table \ref{espectra}.

     The radio emission of the jet lobes was proposed to arise from the interaction of the stellar jet with the surrounding medium. This interaction consists in shocks where a small fraction of the electrons are accelerated to relativistic velocities producing non-thermal emission if a magnetic field is present. In a similar study, 
\citet{vig2018} found negative spectral indices for the knots of the HH 80/81/80N jet steeper at the lowest frequencies observed over this jet (325 and 610 MHz) and was interpreted as thermal free-free contribution at higher frequencies. The lack of significant variations in our derived spectral indices along the 325 MHz-8 GHz range suggest that the contribution of thermal emission is not important at these frequencies. On the other hand, the spectral index found for the central source is positive, consistent with that derived in \citet{garay2003} and \citet{rodriguez2005},  indicating that the emission is predominantly thermal.

As seen in Figure \ref{jet_maps}, the emission in the southern lobe at 325 MHz appears to arise from a position different than the position traced by higher frequency data. We propose two possibilities: first, the southern 325 MHz  knot corresponds to a component aside from the jet. For instance, this knot could be associated with the time variable Source B \citep{rodriguez2005}. Otherwise, the puzzling location of the southern knot at 325 MHz could be representative of non-resolved substructure belonging to the southern lobe of the jet. Although the first possibility cannot be fully discarded, the good fit for the southern knot seen in Figure \ref{spectra} favors the second possibility.  Therefore, in the following discussion we will consider that the southern knot detected at 325 MHz belongs to the jet.

The high angular observations of \citet{rodriguez2005} reveal that the lobes of the jet are composed by several components. The authors derived spectral indexes separately for these components and found variations between them, though most of the indices were negative. The angular resolution of our observations and those of \citet{garay2003} implies that the fluxes of Fig. \ref{spectra} include blended emission from different components and, hence, our spectral indices are representative of the average distribution of the relativistic electron population along the lobe. In this sense, the position of the southern emission peak at 325 MHz suggests that the electrons emitting mostly at these frequency are not concentrated on the southern tip of the shock, contrary to what occurs in the northern lobe.

Non-thermal emission is widely found in extragalactic objects such as quasars and radiogalaxies, that posses a central engine capable of launching relativistic jets. Differences in the emission of the expanding lobes depending on the observed frequency have been reported in extragalactic jets \citep[e.g.,][]{perley1984}. Synchrotron decay due to electron cooling has been invoked to explain the differences observed between frequencies. A common assumption in the study of synchrotron emitting sources is the minimum energy condition that corresponds to equipartition of energy between relativistic particles and magnetic field. Following the prescription of \citet{pacholczyk1970}, we can obtain an estimation of magnetic field in the lobes of the jet under this condition using the expression:

\begin{equation}
\left[\frac{B_\mathrm{eq}}{\mathrm{gauss}}\right] = 5.69 \times 10^{-5} \left[\frac{1+ \chi}{f (\mathrm{sin}\phi)^{3/2}}\left(\frac{\mathrm{arcsec}}{\theta_x \theta_y}\right)\left(\frac{\mathrm{kpc}}{s}\right)S_\mathrm{rad}\right]^{2/7}
\end{equation}

 where $\chi$ is the energy ratio of the heavy particles to electrons for which we adopted 40 as measured in cosmic rays collected near the Earth \citep{simpson1983}, $\phi$ is the angle between the magnetic field and the line of sight (we adopted sin$\phi$ as 0.5), $s$ is the path length through the source in the line of sight that was estimated as $\sqrt{\theta_x\theta_y}$ and converted to linear size accounting for the distance to the source, $\alpha$ is the spectral index, whose values are reported in Table \ref{espectra}, $\theta_x$ and $\theta_y$ are the major and minor axis of the source size (we adopted $11'' \times 4''$, see Table \ref{sources_tot}), $f$ is the filling factor of the source (assumed to be 0.5) and $S_\mathrm{rad}$ is the total flux density integrated from 0.01 to 100 GHz \citep[e.g.][]{miley1980} (typically between 0.04 and 0.05 \jpb$\times$GHz) that is obtained from the measured fluxes and spectral indices from Tables \ref{sources_tot} and \ref{espectra}. From this expression, we obtain a rough estimate for the magnetic field of $\sim$0.5 mG. Despite obtaining a magnetic field value somewhat lower than the obtained by \citet{rodriguez2008}, both values are still consistent within the uncertainties inherent to our estimation and they fall within the range of values estimated in protostellar jets embedded in molecular clouds \citep[0.1-1 mG:][]{carrascogonzalez2010, ainsworth2014, masque2015, vig2018}.      

Following a formalism based on \citet{longair2011}, the cooling time  due to radiative processes for relativistic electrons, $t_\mathrm{c}$, is given by  $t_\mathrm{c}(yr) \sim 2.3 \times 10^4 \nu^{-1/2} B^{-3/2}$, where $\nu$ is our observing frequency in Hz and $B$ is the magnetic field given in Gauss. Using the B value estimated above, we get a $t_\mathrm{c}$ of $(8-10) \times 10^4$ yr for our observed frequency values. 
This range encompasses the  decay time value derived in \citet{rodriguez2008} ($t_\mathrm{d}$ of $9 \times 10^4$ yr). In fact, if we expand the range of  decay time to all observable radio frequencies (up to $\sim$ 40 GHz) we get $\sim10^{4-5}$ yr.  According to the jet velocity \citep[490 \kms,][]{rodriguez2008} we obtain a dynamical age of 280 yr.  The much larger decay time for the synchrotron emission in the shocks makes unlikely the cooling mechanism for electrons as the shock evolves as responsible for the differences in the emission of the jet found between frequency bands.  

Alternatively, different conditions of the associated shock/ambient properties of the southern lobe, with respect the northern lobe could be responsible for the observed discrepancies between frequencies. Contrary to extragalactic jets, protostellar jets dig into a denser inhomogeneous medium. Such density inhomogeneities are inherent to molecular clouds in a wide range of scales \citep[e.g.][]{morata2003}. In this scenario, the jet would impact on small unresolved clumps along the path with a variety of densities. Depending on the density of the ambient,   
the acceleration mechanism for electrons would be strength or inhibited. Then, the discrepancy of the appearance of 325 MHz emission can be explained by collisions with the ambient gas at a given density, which could deplete the low energy electrons from the shock \citep{casillasperez2016} from the tip of the southern lobe. This naturally explains the differences in brightness between the northern and southern lobes. In a similar  recent work, although electron cooling was proposed as a possibility to explain flux variability observed in some knots in the HH 80/81/80N jet \citep{vig2018}, the authors could not rule out associating the emission variability with the fact that the jet passes through an inhomogeneous medium.  Indeed, \citet{rodriguez2008} measured no significant proper motions for the knots and concluded that they are likely part of  the working surfaces of the jet with the ambient medium.  Therefore, the differences in emission between frequency bands and between the northern and southern lobes of the jet could be explained in therms of the interaction of the material of the jet with the medium through it advances, whose inhomogeneities affect the acceleration and depletion mechanisms for the electrons. Observations at mm wavelengths with higher angular resolution than those presented in this work could shed light on this possibility.

\section{Summary}

The very low frequency radio emission (325 MHz and 610 MHz) of the jet associated with \iras\ have been explored. 
Despite detecting clearly the jet at these frequencies, important differences between frequency bands become evident. After accurately align the maps, we found that the knots seen at 325 MHz are not totally coincident with the position of the knots associated with the jet seen at higher frequencies. We have calculated spectral indices by means of the integrated emission of the whole lobes and find consistent results with previous works showing that the non thermal emission is also present down to the observed frequencies.  To explain the discrepant appearance of the jet between frequency bands, we discuss electron cooling in a magnetic field of $\sim0.5$ mG as a possibility. However, the synchrotron decay time is much larger than the dynamical age of the jet. Alternatively, we invoke interaction with an inhomogeneous medium as the most plausible scenario, though high resolution mm observations are required to assess this possibility.

The present work highlights the capability of very low frequency studies towards protostellar jets to explore some of its properties. At present, detailed simulations of the Fokker-Planck equation are being developed in order to extract conclusive results, which will be included in a forthcoming paper.

\acknowledgments

We thank Heinz Andernach for his crucial help with the  astrometry of the maps. We are also grateful to Anabella Araudo for the fruitful comments on synchrotron losses.

\bibliography{IRAS1654}

\clearpage

\input{observations.tex}

\input{sources.tex}

\input{spectra.tex}

\end{document}

%% file: observations.tex
\begin{table}[ht]
\begin{center}
\caption{Details of the GMRT observations\label{observations}}
\vspace{0.5cm}
\begin{tabular}{cccccc}
\hline
\hline
Frequency & Observation & Bandwidth &  \multicolumn{2}{c}{synthesized beam} & $\sigma_\mathrm{rms}^{a}$\\
(MHz)        & date &  (MHz)  &($'' \times ''$) & ($\arcdeg$) & m\jpb  \\
\hline
325   &  2017 Feb 21  &  32   &  24.2    $\times $     5.8  &   12.2  &  0.6   \\
610   &  2017 Feb 28  &  32   &  12.4    $ \times $    3.3  &   15.6  &  0.3   \\
\hline
\end{tabular}
\end{center}
$^{a}$ {Measured around the phase reference center, where the jet is found}\\
\end{table}

%% file: sources.tex
\begin{landscape}
\begin{table}[ht]
\footnotesize
\begin{center}
\vspace{-1.7cm}
\caption{Parameters of the sources in IRAS16547-4247$^{a}$\label{sources_tot}}
\begin{tabular}{p{0.9cm}cccccccccc}
\hline
\hline
& \multicolumn{2}{c}{Coordinates$^{b}$}    &  $\theta_M$$^{b,c}$  &  $\theta_m$$^{b,c}$  &  P.A.$^{b,c}$  & $S_\mathrm{610 MHz}$  & $I_\mathrm{610 MHz}^{Peak}$&$S_\mathrm{325 MHz}$  & $I_\mathrm{325 MHz}^{Peak}$\\
Knot &  $\alpha$~(J2000) & $\delta$~(J2000)   &   ($''$)  & ($''$)    &   (deg.) &  (mJy) & (mJy bm$^{-1}$)&  (mJy) & (mJy bm$^{-1}$)\\
 \hline
North & 16:58:16.926 & -42:51:55.26  & $ 11.0 \pm 2.4$ & $ 3.9 \pm 0.6 $ & $ 7.4 \pm $ 7.5 & $ 7.4 \pm 1.6$ & $ 3.5 \pm 0.5$         & $ 7.3 \pm 1.6$ & $ 5.3 \pm 0.6$ \\
Central & 16:58:17.195 & -42:52:04.64 & -- & --  & --  & $ 2.7 \pm 1.2$ & $ 2.4 \pm 0.5$ 	& 1.3$^{d}$ & -- \\
South$^{e}$ & 16:58:17.437 & -42:52:12.63 & $ 11.3 \pm 5.2 $ & $ 4.1 \pm 2.1 $ & $ 5.9 \pm $ 28.6 & $ 3.9 \pm 1.5$ & $ 1.7 \pm 0.5$		& $ 5.4 \pm 1.7$ & $ 4.3 \pm 0.7$ \\
 \hline
\end{tabular}
\end{center}
$^{a}$ {Obtained from simultaneous Gaussian fits (see text).}\\
$^{b}$ {Results from the fits at 610 MHz map.}\\
$^{c}$ {Deconvolved source size, except for the C source that is unresolved.}\\
$^{d}$ {The value corresponds from an estimation based on the liner regression presented in the middle panel of Fig. \ref{spectra} that approximately coincides with the upper limit reported at 325 MHz (see also Subsection 3.2).}\\
$^{e}$ {Knots at are not coincident between both frequency bands}\\
\end{table}
\end{landscape}

%% file: spectra.tex
\begin{table}[ht]
\footnotesize
\begin{center}
\vspace{-1.7cm}
\caption{Spectral indexes of the sources in IRAS16547-4247$^{a}$\label{espectra}}
\begin{tabular}{ccc}
\hline
\hline
Source &$\alpha^{a}$ & $r^2$ $^{b}$ \\
 \hline
Northern lobe & $-0.45 \pm 0.10$ &0.81\\
Center & $0.44 \pm 0.08$ &0.87 \\
Southern lobe & $-0.48 \pm 0.04$ &0.97 \\
 \hline
\end{tabular}
\end{center}
$^{a}$ {Spectral index obtained including data points from \cite{garay2003} and the present work. We excluded some data points (see text).}\\
$^{b}$ {Square of correlation coefficient.}\\
\end{table}